\documentclass[prl,twocolumn,amsmath,amssymb]{revtex4-1}
\usepackage{graphicx}
\usepackage{dcolumn}
\usepackage{bm}

\begin{document}
\title{Extraordinary electron transmission through a periodic array of quantum dots}
\author{L. S. Petrosyan,$^{1,2}$  A. S. Kirakosyan,$^2$ and T. V. Shahbazyan$^2$} 
\affiliation{
$^1$Department of Medical Physics, Yerevan State Medical University, 2 Koryun Street, Yerevan, 0025, Armenia\\
$^2$Department of Physics, Jackson State University, Jackson, Mississippi
39217 USA
}
\begin{abstract}
We study electron transmission through a periodic array of quantum dots (QD) sandwiched between doped semiconductor leads. When the Fermi wavelength of tunneling electron exceeds the array lattice constant, the off-resonant per QD conductance is enhanced by several orders of magnitude relative to the single-QD conductance. The physical mechanism  of the enhancement is delocalization of a small fraction of system eigenstates caused by coherent coupling of QDs via the electron continuum in the leads.
\end{abstract}
\maketitle

Interference effects in electron transmission through localized states such as, e.g., semiconductor quantum dots (QD) have been among the highlights in coherent  transport studies  \cite{yacoby-prl95,shuster-nature97}. The phase acquired by an electron tunneling along multiple pathways provided by QDs placed between doped semiconductor leads can cause striking features in tunneling current \cite{buks-nature98}. The simplest realization of coherent transport is provided by \textit{two} QDs independently coupled to the leads at weak (or absent) interdot tunneling \cite{shahbazyan-prb94}.  The conductance of a double-QD system in a magnetic field  exhibits Aharonov-Bohm oscillations \cite{blick-prl01,ensslin-prl06,hatano-prl11} as a function of magnetic flux penetrating the area bound  by tunneling paths \cite{shahbazyan-prb94,loss-prl00,gefen-prl01}. At zero field, the coherence is controlled by QDs' coupling via the electron continuum in the leads \cite{shahbazyan-prb94,brandes-pr05}. If QDs separation, $a$, is comparable to electron's Fermi wavelength, $\lambda_{F}$, then electron transmission is mediated by the system eigenstates rather than individual QDs, leading to conductance peak narrowing or Fano-like lineshapes \cite{shahbazyan-prb94,kubala-prb02,brandes-prb03,orellana-prb03}. A close optical analogy for coherent transport through QDs is the Dicke superradiance of two excited atoms \cite{dicke-pr54}; QDs' coupling via the electron continuum is similar to radiative coupling between the atoms \cite{shahbazyan-prb94,brandes-pr05}.

\begin{figure}[tb]
\begin{center}
\includegraphics[width=0.95\columnwidth]{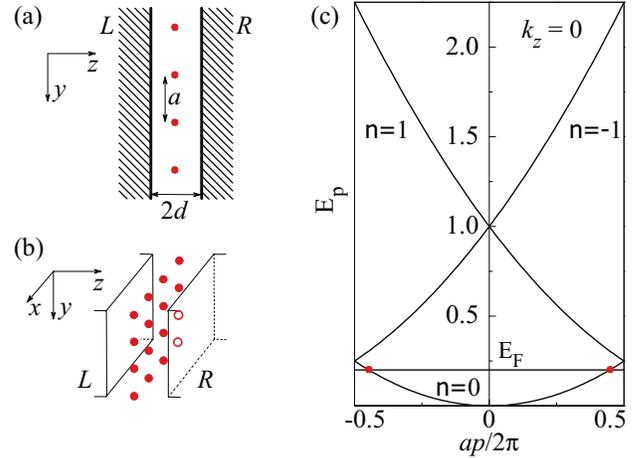}
\caption{\label{fig:picture}
(Color online) Schematics of (a) 1D periodic chain of QDs between 2D leads (1DQD-2DEG), (b) 2D lattice of QDs between 3D leads (2DQD-3DEG), and (c) lowest EG bands ($g=0,\pm 2\pi/a$)  in $k_{z}=0$ plane in units of $E_{a}=\pi^{2}/2ma^{2}$; vHS is located at the crossings of bands and Fermi level.}
\end{center}
\end{figure}

In this Letter, we study electron transport through a periodic array of localized states, e.g., a QD lattice with period $a$ separated from the electron gas (EG) in the leads by tunneling barriers. We consider a linear chain of QDs  situated between two-dimensional (2D) leads (1DQD-2DEG) and a square lattice of QDs sandwiched between 3D leads (2DQD-3DEG), as illustrated in Fig.~\ref{fig:picture}.  We demonstrate that in the subwavelength regime, i.e., $a/\lambda_{F}\lesssim  1$, the \textit{off-resonant} per QD conductance is strongly enhanced  relative to single-QD one. We trace the origin of this effect to \textit{delocalization} of a small fraction of system's eigenstates which mediate electron transport across tunneling barriers. 

The novel coherent transport effect predicted  here bears similarity to extraordinary optical transmission (EOT) of light through a  metal film perforated with a periodic  array of nanoholes \cite{ebbesen-nature98}. Suggested  EOT mechanisms involve surface plasmon-polaritons \cite{ebbesen-nature98,ebbesen-prb98,ebbesen-prl01}, dynamical diffraction \cite{treacy-prb02}, light's coupling to waveguide resonances \cite{lalanne-prl02}, and composite diffracted evanescent wave \cite{lezec-oe04}. However, the physical origin of EOT remains under active discussion \cite{lezec-naturephys06,abajo-rmp07,lalanne-nature08}. A similar effect  was recently reported for a transmission of sound waves \cite{christensen-naturephys07,lu-prl07,estrada-prl08}. Here we describe a related  phenomenon in electron transport through a QD array. Despite differences with EOT, the generic effect of enhanced transmission persists. In this case, however, we can pinpoint the  mechanism of extraordinary \textit{electron} transmission (EET).

Indeed, consider the electronic states in the system comprised of QD lattice (1D or 2D) tunnel-coupled to EGs in the leads (2D or 3D) without direct tunneling between QDs (see Fig.~\ref{fig:picture}). The periodic potential of QD lattice splits the EG momentum space $({\bm k},k_{z})$ into bands in \textit{transverse} direction $({\bm p}+{\bm g},k_{z})$, where ${\bm p}$ lies in the first Brillouin zone (BZ), ${\bm g}$ are reciprocal lattice vectors, and $k_{z}$ is the momentum along tunneling direction. At Fermi level, the electronic states in band ${\bm g}$ with energy $E_{F}=({\bm p}+{\bm g})^{2}/2m+k_{z}^{2}/2m$ are either propagating for $|{\bm p}+{\bm g}|< k_{F}$ (real $k_{z}$), or evanescent for $|{\bm p}+{\bm g}|> k_{F}$ (imaginary $k_{z}$), where $k_{F}=\sqrt{2mE_{F}}=2\pi/\lambda_{F}$ and $m$ are the electron Fermi momentum and mass, respectively, (we set $\hbar =1$).  At the same time, the coupling between QDs via EG gives rise to localized states' band (LSB)  whose eigenstates are characterized by transverse quantum number ${\bm p}$ and are confined along $z$ direction. The LSB eigenstates' dispersion $E_{\bm p}$ comes from coherent coupling between QDs' via \textit{evanescent} states of EG, while eigenstates' decay into EG \textit{propagating} states determines their lifetime. Importantly, the conservation of ${\bm p}$  \textit{reduces} the phase space for tunneling to that of a discrete set of 1DEGs (enumerated by ${\bm g}$) with dispersion $k_{z}^{2}/2m$. A transition between an LSB eigenstate ${\bm p}$ and EG  is then determined by the 1DEG partial density of states (DOS) which has the van Hove singularity (vHS) at $|k_{z}|\equiv \sqrt{k_{F}^{2} -({\bm p}+{\bm g})^{2}}=0$ (see Fig.~\ref{fig:picture}). An electron in LSB  eigenstate with ${\bm p}$ near vHS can escape to the leads and hence is delocalized across tunneling barriers.  

Specifically, a single-QD resonant tunneling conductance through level $E_{0}$  is given by standard expression
\begin{align}
\label{cond-single}
G_{0}=\dfrac{e^{2}}{\pi}\dfrac{\Gamma_{L}\Gamma_{R}}{\bigl(E_{F}- E_{0}\bigr)^{2}+\frac{1}{4}(\Gamma_{L}+\Gamma_{R})^{2}},
\end{align}
where $\Gamma_{L,R}$ are decay rates to left or right leads.  In the symmetric case $\Gamma_{L}=\Gamma_{R}=\Gamma$, $G_{0}=e^{2}/\pi $ at resonance $E_{F}=E_{0}$, while the off-resonant $G_{0}$ is reduced by  factor $\left (\Gamma/\delta E\right )^{2}\ll 1$, where $\delta E=E_{F}-E_{0}$ is the Fermi level detuning. We show that the low-energy  per QD array conductance $G$ is enhanced relative to $G_{0}$ by the factor
\begin{equation}
\label{enh}
\frac{G}{G_{0}}\approx \frac{2\pi}{(ak_{F})^{2}}\ln \left (\frac{|\delta E|}{ \gamma}\,ak_{F}\right ),
\end{equation}
$\gamma= (2\pi/ak_{F}) \Gamma$ being $k_{F}$-independent, which can reach several orders of magnitude. Equation (\ref{enh}) holds in the 2DQD-3DEG case; in  the 1DQD-2DEG case the prefactor is replaced by $4/\pi ak_{F}$ and $\gamma=2\Gamma$. In both cases, the prefactor reflects QDs' number in the volume with linear size $\lambda_{F}$ and results from the coherent coupling between QDs \cite{shahbazyan-prb98}, while the logarithmic factor describes delocalized states' contribution. For asymmetric leads, enhancement is reduced due to different vHS locations in ${\bm p}$-space for left and right leads.

\textit{Theory}.--- We start with tunneling Hamiltonian for a square lattice of $N$ QDs with in-plane coordinates ${\bm r}_{j}$ separated by potential barriers from 3DEGs in the leads
\begin{equation}
\label{h}
H=\sum_{j} E_{0}c_{j}^{\dagger}c_{j}+ 
\sum_{\nu\alpha}{\cal E}_{\nu}^{\alpha}c_{\nu\alpha}^{\dagger}c_{\nu\alpha}+
\sum_{\nu\alpha j} \left (V_{j\nu}^{\alpha} c_{j}^{\dagger}c_{\nu\alpha}+\text{H.c.}\right ),
\end{equation}
where  $c_{j}^{\dagger}$, $c_{j}$ and $ E_{0}$ are, respectively, creation and annihilation operators and energies for localized states, $c_{\nu\alpha}^{\dagger}$, $c_{\nu\alpha}$, and ${\cal E}_{\nu}^{\alpha}$ are those for EG states in the leads ($\alpha=L,R$), and $V_{\nu j}^{\alpha}$ is transition matrix element between localized and EG states. We assume the barrier thickness smaller than the lattice period and neglect the direct tunneling between QDs. The zero-temperature conductance normalized \textit{per} QD is given by  \cite{brandes-pr05}
\begin{equation}
\label{cond}
G=\frac{e^{2}}{\pi\hbar}\frac{\rm Tr}{N}\left(\hat{\Gamma}^{R}\frac{1}{E_{F}-E_{0}-\hat{\Sigma}} \,\hat{\Gamma}^{L}\frac{1}{E_{F}-E_{0}-\hat{\Sigma}^{\dagger}}\right),
\end{equation}
where $\Sigma_{ij}=\Sigma_{ij}^{L}+\Sigma_{ij}^{R}$ is the self-energy matrix, 
\begin{equation}
\label{self-matrix}
\Sigma_{ij}^{\alpha} =\sum_{ \nu}\frac{V_{i\nu}^{\alpha}V_{\nu j}^{\alpha}}{E_{F} -{\cal E}_{\nu}^{\alpha}+i0} = \Delta_{ij}^{\alpha}-\frac{i}{2} \Gamma_{ij}^{\alpha},
\end{equation}
whose principal and singular parts determine the energy matrix $\Delta_{ij}^{\alpha}$ and the decay matrix $\Gamma_{ij}^{\alpha}$, respectively, and the trace is taken over lattice sites. The matrix element can be presented as $V_{j\nu}^{\alpha}=(AL_{z})^{-1/2}e^{i{\bm k}\cdot {\bm r}_{j}}t_{\alpha}$, where ${\bm k}$ and ${\bm r}_{j}$ are the \textit{in-plane} (transverse) momentum and coordinate, respectively,  $t_{\alpha}$ is  the tunneling amplitude between QD and the lead $\alpha$ (we assume that the barrier is sufficiently high and neglect its dependence on energy), and $A=Na^{2}$ and $ L_{z}$ are normalization area and length, respectively. The self-energy is expressed via the electron Green function as $\Sigma_{ij}^{\alpha} =t_{\alpha}^{2}G_{\alpha}({\bm r}_{i},d_{\alpha};{\bm r}_{j},d_{\alpha})$, $d_{\alpha}$ being the  barrier thickness.

The coupling between QD lattice and EG gives rise to LSB whose eigenstates are characterized by \textit{in-plane} quasimomentum ${\bm p}$ lying in the first BZ. At the same time, QDs' periodic potential splits 3DEG momentum space (${\bm k},k_{z}$) into bands (${\bm p}+{\bm g},k_{z}$), where ${\bm g}=(2\pi m/a,2\pi n/a)$, $m$ and $n$ being integers. The LSB  spectrum is obtained via Fourier transform of  Eq.~(\ref{self-matrix}), 
\begin{equation}
\label{fourier}
\Sigma_{ij}^{\alpha}=\frac{1}{N}\sum_{\bm p}e^{i{\bm p}\cdot ({\bm r}_{j}-{\bm r}_{j})}\Sigma_{\bm p}^{\alpha},~~
\Sigma_{\bm p}^{\alpha}=\frac{t_{\alpha}^{2}}{a^{2}}\sum_{\bm g}\int \frac{dk_{z}}{2\pi}G^{\alpha}_{{\bm g}+{\bm p},k_{z}},
\end{equation}
where  $\Sigma_{\bm p}^{\alpha}$  and  $G^{\alpha}_{{\bm k},k_{z}}=\left [E_{F}-(k^{2}+k_{z}^{2})/2m_{\alpha}\right ]^{-1}$ are, respectively, the LSB self-energy and the Fourier of electron Green function, $m_{\alpha}$ being the electron mass. After $k_{z}$-integration, the LSB self-energy takes the form
\begin{equation}
\label{self}
\Sigma_{\bm p}^{\alpha}=\Delta_{\bm p}^{\alpha}-\frac{i}{2}\Gamma_{\bm p}^{\alpha}=-\frac{i}{2a}\sum_{\bm g}\frac{\gamma_{\alpha}}{\sqrt{\left (k_{F}^{\alpha}\right )^{2}-({\bm g}+{\bm p})^{2}}},
\end{equation}
where $\gamma_{\alpha}=2m_{\alpha}t_{\alpha}^{2}/a$, and $k_{F}^{\alpha}=\sqrt{2m_{\alpha}E_{F}}$.  Note that for a given ${\bm p}$, each band contributes either to $\Delta_{\bm p}^{\alpha}$ or to $\Gamma_{\bm p}^{\alpha}$ depending on whether $|{\bm g}+{\bm p}|>k_{F}^{\alpha}$ or $|{\bm g}+{\bm p}|<k_{F}^{\alpha}$. Thus, QDs' coupling via evanescent states of EG gives rise to the LSB dispersion $E_{\bm p}=E_{0}+\Delta_{\bm p}^{L}+\Delta_{\bm p}^{R}$ (note that $\Delta_{\bm p}^{\alpha}<0$), while transitions between LSB and propagating states of EG determine the decay rate $\Gamma_{\bm p}^{\alpha}$. The LSB self-energy Eq.~(\ref{self}) diverges at $|{\bm g}+{\bm p}|=k_{F}^{\alpha}$, i.e., for EG states propagating along the interface, due to vHS in the 1DEG DOS at $k_{z}=0$. The latter is caused by conservation of ${\bm p}$ which reduces the phase space for tunneling to that of 1DEG with dispersion $k_{z}^{2}/2m$. An LSB eigenstate with ${\bm p}$ near vHS has a high decay rate to the EG and is, in effect,  delocalized across the barriers.  As we show below, these states mediate EET through a QD lattice.

Using Fourier transform (\ref{fourier}), the conductance (\ref{cond}) can be recast in terms of system's eigenstates as
\begin{equation}
\label{cond2}
G=\dfrac{e^{2}}{\pi}
\left (\frac{a}{2\pi}\right )^{2}\int d{\bm p}\,
\frac{\Gamma_{\bm p}^{L}\Gamma_{\bm p}^{R}}{\left (E_{F}-E_{\bm p}\right ) ^{2}+\frac{1}{4}\left (\Gamma_{\bm p}^{L}+\Gamma_{\bm p}^{R}\right )^{2}},
\end{equation}
where ${\bm p}$-integral is taken over 2D BZ.  Thus, $G$ is a sum of partial conductances for each ${\bm p}$: $G=\left (\frac{a}{2\pi}\right )^{2}\!\int \!d{\bm p}\,G_{\bm p}$, where $G_{\bm p}$ has the form (\ref{cond-single}) with $E_{0}$ and $\Gamma_{\alpha}$ replaced by $E_{\bm p}$ and $\Gamma_{\bm p}^{\alpha}$, respectively. In 1DQD-2DEG case, ${\bm p}$ and  ${\bm g}$ in Eq.~(\ref{self}) are 1D vectors (and $\gamma_{a}=2m_{\alpha}t_{\alpha}^{2}$), while integration measure in Eq.~(\ref{cond2}) extends over 1D BZ. 

For a given $E_{F}$, only EG bands with propagating states, i.e., those with $|{\bm g}+{\bm p}|<k_{F}^{\alpha}$, contribute to  $\Gamma_{\bm p}^{\alpha}$ and hence provide transmission channels. If $E_{F}$ is larger than the geometric energy scale  $E_{a}=\pi^{2}/2ma^{2}$, i.e., the coherent coupling between QDs is weak ($ak_{F}\gg 1$), then a large number of bands  contribute to $G$, and replacing the sum over ${\bm g}$ in Eq.~(\ref{self}) with integral one recovers the single-QD decay rate $\Gamma_{\bm p}^{\alpha}= \Gamma_{\alpha}$ and conductance (\ref{cond-single}) for uncoupled QDs. However, for $E_{F}\lesssim E_{a}$, i.e., strong coherent coupling ($ak_{F}\lesssim 1$), only lowest EG bands contribute to decay rate $\Gamma_{\bm p}^{\alpha}$, and $G$ is determined by its behavior near vHS.

\begin{figure}[tb]
\begin{center}
\includegraphics[width=0.93\columnwidth]{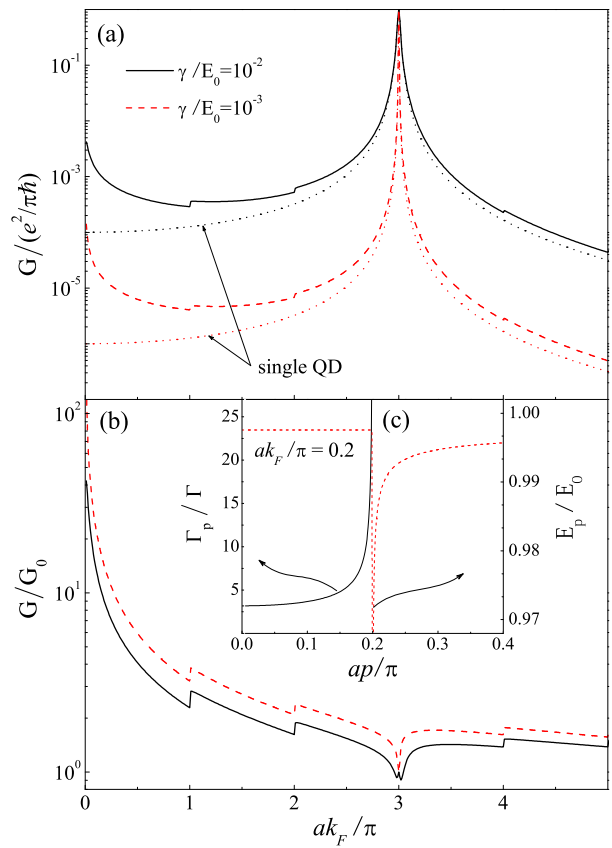}
\caption{\label{fig:2D1D}
(Color online) 1DQD-2DEG normalized conductance (a) and enhancement factor (b) vs. EG Fermi momentum are shown for symmetric case at $E_{0}/E_{a}=9.0$ for two values of $\gamma/E_{0}$; (c) vHS in LSB energy spectrum at $ak_{F}/\pi=0.2$. }
\end{center}
\end{figure}

\begin{figure}[tb]
\begin{center}
\includegraphics[width=0.93\columnwidth]{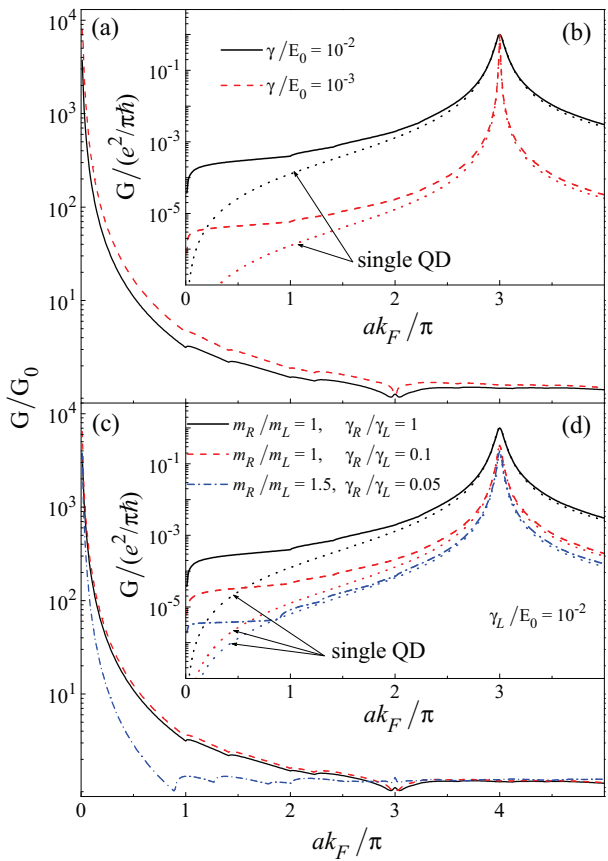}
\caption{\label{fig:2D3D}
(Color online) 2DQD-3DEG enhancement factor (a) and normalized conductance (b) vs. EG Fermi momentum are shown for symmetric case at $E_{0}/E_{a}=9.0$ for two values of $\gamma/E_{0}$; (c) and (d) -- symmetric case is compared to asymmetric case for $\gamma_{L}/E_{0}=10^{-2}$.}
\end{center}
\end{figure}

In Fig.~\ref{fig:2D1D} we show the conductance and enhancement factor for symmetric 1DQD-2DEG case ($\Gamma_{\bm p}^{\alpha}=\Gamma_{\bm p}$, $\gamma_{\alpha}=\gamma$, $k_{F}^{\alpha}=k_{F}$). Numerical calculations were carried using Eqs. (\ref{self}) and (\ref{cond2}) with QD lattice period chosen to set $E_{0}=9E_{a}$, corresponding to $ak_{F}=3\pi$ at resonance, and ${\bm g}$ with $m,n$ up to 50 included. For $ak_{F}/\pi < 1$, only ${\bm g}=0$ band contributes to $\Gamma_{\bm p}$;  $\Gamma_{\bm p}\neq 0$ only below vHS ($p<k_{F}$), while $E_{\bm p}$ is affected by vHS for $p>k_{F}$ [see Fig.~\ref{fig:2D1D}(c)]. Note that higher bands' contribution leads to energy renormalization by a constant that causes apparent deviation of $E_{\bm p}/E_{0}$ from unity below vHS (for the given maximal $g$); in the following, we absorb this constant into $E_{0}$. Note also that $E_{\bm p}/E_{0}$ is suppressed relative to $\Gamma_{\bm p}/\Gamma$ by  small factor $\Gamma/E_{0}\ll 1$,  and so the LSB dispersion $E_{\bm p}$ has no significant effect on $G$. In fact, the LSB DOS (not shown here) is nearly indistinguishable from that of a single QD, indicating that vHS affects only a small fraction of LSB. However, it is this small fraction of LSB eigenstates ${\bm p}$ with large decay rates, $\Gamma_{\bm p}\gtrsim |\delta E|$, that causes dramatic conductance enhancement on the low energy side [see Fig.~\ref{fig:2D1D}(a)]. Indeed, the partial conductance of these \textit{off-resonant} states is close to the maximal (ballistic) value [$G_{\bm p}\sim e^{2}/\pi$ in Eq.~(\ref{cond2})] indicating that these states are, in effect, delocalized. For $ak_{F}/\pi <1$, $G$ can be evaluated  as (in 2DQD-3DEG case)
\begin{align}
\label{cond3}
G&=\dfrac{e^{2}}{\pi}\frac{a^{2}}{2\pi}\int_{0}^{k_{F}}\frac{dpp\left( \gamma/a\right)^{2}\left(k_{F}^{2}-p^{2}\right)^{-1}}{\left (\delta E\right )^{2}+\left( \gamma/a\right)^{2}\left(k_{F}^{2}-p^{2}\right)^{-1}}
\nonumber\\
&\approx
\dfrac{e^{2}}{\pi}\frac{1}{2\pi}\left (\frac{\gamma}{\delta E}\right )^{2}\ln\left (\frac{|\delta E| }{\gamma}\, ak_{F}\right ),
\end{align}
where chief contribution to the integral comes from the region $(p-k_{F})/k_{F}\sim \left(\gamma/\delta E a k_{F}\right )^{2}\ll 1$ constituting  $\sim \left (\gamma/\delta E\right )^{2}$ fraction of 2D BZ, while the log factor comes from the slow (square root) divergence of vHS and reflects delocalized states' distribution in energy. In 1DQD-2DEG case, $G$ differs from the result (\ref{cond3}) by factor $2/ak_{F}$. The sharp features in $k_{F}$-dependence of $G$ occur when the Fermi level crosses EG band edges. The enhancement factor $G/G_{0}$ is now determined by taking into account the relations $\gamma=2\Gamma$ for 1DQD-2DEG and $\gamma=(2\pi/ak_{F})\Gamma$ for 2DQD-3DEG case, yielding $G/G_{0}=(4/\pi a k_{F})\ln\left (ak_{F}|\delta E|/\gamma \right )$ in the former case and Eq.~(\ref{enh}) in the latter case. In the region $\gamma/|\delta E|\ll k_{F}a\ll 1$, both logarithm and its prefactor contribute to overall enhancement that reaches 2 orders of magnitude in 1DQD-2DEG case [see Fig.~\ref{fig:2D1D}(b)]; however, the $(ak_{F})^{-2}$ prefactor dependence in Eq.~(\ref{enh}) leads to greater enhancement in 2DQD-3DEG case, as shown in Figs.~\ref{fig:2D3D}(a) and \ref{fig:2D3D}(b). 

Breaking the left-right symmetry significantly alters the conductance; however, a distinction must be made in the way it is broken [see Fig.~\ref{fig:2D3D}(c) and \ref{fig:2D3D}(d)]. For example, a smaller $\Gamma_{\bm p}^{R}$ due to higher tunneling barrier between QDs and the right lead causes overall reduction of both array and single-QD conductances; however the enhancement factor $G/G_{0}$ stays essentially unchanged. In contrast, an EG asymmetry, e.g., due to different effective masses ($m_{L}\neq m_{R}$), drastically lowers the enhancement. The reason is that  for $k_{F}^{L}\neq k_{F}^{R}$, the LSB eigenstates have different decay rates to left and right EGs due to different respective vHS locations in the ${\bm p}$-space. Here the enhancement factor is easily evaluated  as $G/G_{0}\approx 2\pi /(ak_{F})^{2}\sqrt{m_{R}/m_{L}-1}$ for $(m_{R}/m_{L}-1)\sim 1$, i.e., it is suppressed  by log factor relative to Eq.~(\ref{enh}).

Finally, consider the effect of a finite temperature, $T$, and of a weak disorder due to QDs' size and position distributions. The disorder can be characterized by inhomogeneous width of QDs energies, $\gamma_{0}\ll \delta E$, and deviation of lattice constant from its average value,  $\delta a \ll a$. The former lowers the resonant transmission but does not affect the off-resonant one dominated by states with $\Gamma_{\bm p}\sim \delta E$. Similarly, finite $T$ broadens the resonance, but the states with $\Gamma_{\bm p} \gg T$ remain unaffected. However, even a weak nonperiodicity  breaks down exact conservation of  ${\bm p}$ and hence smears out vHS thus reducing the fraction of delocalized states. It is easy to show that for $\delta a/a\gtrsim \left (\gamma/ \delta E \right )^{2}$, Eq.~(\ref{enh}) still holds, but with the logarithm argument replaced by $\sqrt{a/\delta a}$.

In conclusion, there is an apparent similarity between EET through a QD lattice and  EOT picture based on the coupling of light to waveguide modes in a perforated metal film \cite{lalanne-prl02}. Note, however, that despite strong enhancement (up to $10^{4}$ in Fig.~\ref{fig:2D3D}), the off-resonant conductivity is still much lower than its resonance value, in contrast to EOT. The reason can be traced to exponentially small barriers' transparencies which render the fraction of delocalized states to be exponentially small as well. On the other hand, EOT represents a series of well defined peaks while EET through a QD lattice reveals itself in elevated plateau on the conductance low-energy side.  These and other issues will be addressed by us elsewhere. 

This work was supported by the NSF under Grants No. DMR-0906945 and No. HRD-0833178, and by the EPSCOR program.

\end{document}